\documentclass[11pt]{cernrep}
\usepackage{graphicx,epsf,epsfig}
\usepackage{here}
\newcommand{\gsim}{\buildrel > \over {_\sim}}

\begin{document}

\title{INCLUSIVE HEAVY-QUARKONIUM PRODUCTION IN HADRONIC COLLISIONS\footnote
{~~To appear in the CERN Yellow Report for the 2001--2002 CERN workshop
Hard Probes in Heavy Ion Collisions at the LHC.}}
\author{Geoffrey T.~Bodwin$^1$, Jungil Lee$^1$, and Ramona Vogt$^{2,3}$}
\institute{$^1$High Energy Physics Division, Argonne National Laboratory,
Argonne, IL 60439, USA\\
$^2$Nuclear Science Division, Lawrence Berkeley National 
Laboratory, Berkeley, CA 94720, USA\\
$^3$Physics Department, University of California at Davis, 
Davis, CA 95616, USA}
\maketitle
\begin{abstract}
We discuss the theory of inclusive production of heavy quarkonium, the
comparison between theory and experiment, and the expected nuclear
effects in cold nuclear targets. We also present predictions, based on
Nonrelativistic QCD (NRQCD), for quarkonium production cross sections in
the LHC energy range. We find that nuclear effects in the production
cross sections are largely independent of the sizes of the NRQCD matrix
elements, the charmonium state that is produced and, in the color-octet
case, largely independent of the partonic subprocess that produces the
heavy quark-antiquark pair.
\end{abstract}

\section{A Review of NRQCD
}
\label{quarkon.nrqcd1}

\subsection{The NRQCD Factorization Method}\label{NRQCD-factorization}%

In both heavy-quarkonium decays and hard-scattering production, large
energy-momentum scales appear. The heavy-quark mass $m_Q$ is much larger than
$\Lambda_{\rm QCD}$, and, in the case of production, the transverse
momentum $p_T$ can be much larger than $\Lambda_{\rm QCD}$ as well. 
Thus, the associated values of $\alpha_s$
are much less than one: $\alpha_s(m_c)\approx 0.25$ and
$\alpha_s(m_b)\approx 0.18$. Therefore, one might hope that it would be
possible to calculate the rates for heavy quarkonium production and decay 
accurately in perturbation theory. However, there are clearly
low-momentum, nonperturbative effects associated with the dynamics of
the quarkonium bound state that invalidate the direct application of 
perturbation theory. 

In order to make use of perturbative methods, one must first separate
the short-distance/high-momentum, perturbative effects from the
long-distance/low-momentum, nonperturbative effects---a process which is
known as ``factorization.'' One convenient way to carry out this
separation is through the use of the effective field theory
Nonrelativistic QCD (NRQCD)
\cite{Caswell:1985ui,Thacker:1990bm,Bodwin:1994jh}. NRQCD reproduces
full QCD accurately at momentum scales of order $m_Qv$ and smaller,
where $v$ is heavy-quark velocity in the bound state in the
center-of-mass (CM) frame, with $v^2\approx 0.3$ for charmonium and
$v^2\approx 0.1$ for bottomonium. Virtual processes involving momentum
scales of order $m_Q$ and larger can affect the lower-momentum
processes.  Their effects are taken into account through the
short-distance coefficients of the operators that appear in the NRQCD
action.

Because $Q\overline Q$ production occurs at momentum scales of order $m_Q$ or
larger, it manifests itself in NRQCD through contact interactions. As a
result, the quarkonium production cross section can be written as a sum
of the products of NRQCD matrix elements and short-distance coefficients:
\begin{equation}
\sigma(H)=\sum_n {F_n(\Lambda)\over m_Q^{d_n-4}}\langle 0|
{\cal O}_n^H(\Lambda)|0\rangle \, \, .
\label{prod-fact}
\end{equation}
Here, $H$ is the quarkonium state, $\Lambda$ is the ultraviolet cutoff of
the effective theory, the $F_n$ are short-distance coefficients, and the 
${\cal O}_n^H$ are four-fermion operators, whose mass dimensions are 
$d_n$. A formula similar to Eq.~(\ref{prod-fact}) exists for the inclusive
quarkonium annihilation rate \cite{Bodwin:1994jh}.

The short-distance coefficients $F_n(\Lambda)$ are essentially the
process-dependent partonic cross sections to make a $Q\overline Q$ pair. 
The $Q\overline Q$ pair can be
produced in a color-singlet state or in a color-octet state. The
short-distance coefficients are determined by matching the square of
the production amplitude in NRQCD to full QCD. Because the 
$Q\overline Q$ production scale is of order $m_Q$ or greater, 
this matching can be carried out in perturbation theory.

The four-fermion operators in Eq.~(\ref{prod-fact}) create a $Q\overline Q $
pair, project it onto an intermediate state that consists of a heavy
quarkonium plus anything, and then annihilate the $Q\overline Q$ pair. The
vacuum matrix element of such an operator is the probability for a
$Q\overline Q$ pair to form a quarkonium plus anything. These matrix elements
are somewhat analogous to parton fragmentation functions. They contain
all of the nonperturbative physics that is associated with evolution of the
$Q\overline Q$ pair into a quarkonium state.

Both color-singlet and color-octet four-fermion operators appear in
Eq.~(\ref{prod-fact}). They correspond, respectively, to the evolution of
a $Q\overline Q$ pair in a relative color-singlet state or a relative
color-octet state into a color-singlet quarkonium. If we drop all of the
color-octet contributions in Eq.~(\ref{prod-fact}), then we have the
color-singlet model \cite{Schuler:1994hy}. In contrast, NRQCD is not a
model, but a rigorous consequence of QCD in the limit $\Lambda_{\rm
QCD}/m_Q\rightarrow 0$.

The NRQCD decay matrix elements can be calculated in lattice simulations
\cite{Bodwin:1993wf,Bodwin:2001mk} or determined from phenomenology.
However, at present, the production matrix elements must be obtained
phenomenologically, as it is not yet known how to formulate the
calculation of production matrix elements in lattice simulations. In
general, the production matrix elements are different from the decay
matrix elements. However, in the color-singlet case, the production and
decay matrix elements can be related through the vacuum-saturation
approximation, up to corrections of relative order $v^4$
\cite{Bodwin:1994jh}.

An important property of the matrix elements, which greatly increases
the predictive power of NRQCD, is the fact that they are universal, {\it
i.e.}, process independent. NRQCD $v$-power-counting rules 
organize the sum over operators in Eq.~(\ref{prod-fact}) as an
expansion in powers of $v$. Through a given order in $v$, only a limited
number of operator matrix elements contribute. Furthermore, at leading
order in $v$, there are simplifying relations between operator matrix
elements, such as the heavy-quark spin symmetry \cite{Bodwin:1994jh} and the
vacuum-saturation approximation \cite{Bodwin:1994jh}, that reduce the number of
independent phenomenological parameters.  In contrast, the CEM ignores
the hierarchy of matrix elements in the $v$ expansion.

The proof of the factorization formula (\ref{prod-fact}) relies both on
NRQCD and on the all-orders perturbative machinery for proving
hard-scattering factorization.  A detailed proof does not yet exist, but
work is in progress \cite{qiu-sterman}. Corrections to the
hard-scattering part of the factorization are thought to be of order
$(m_Q v/p_T)^2$, not $(m_Q/p_T)^2$, in the unpolarized case and of order
$m_Q v/p_T$, not $m_Q/p_T$, in the polarized case. It is not known if there
is a factorization formula at low $p_T$ or for the $p_T$-integrated
cross section. The presence of soft gluons in the quarkonium 
binding process makes the application of the standard factorization 
techniques problematic at low $p_T$.

In the decay case, the color-octet matrix elements can be
interpreted as the probability to find the quarkonium in a Fock state
consisting of a $Q\overline Q$ pair plus some gluons. It is a common
misconception that color-octet production proceeds, like color-octet
decay, through a higher Fock state. However, in color-octet production,
the gluons that neutralize the color are in the final state, not the
initial state. There {\it is} a higher-Fock-state process, but it
requires the production of gluons that are nearly collinear to the
$Q\overline Q$ pair, and it is, therefore, suppressed by additional
powers of $v$.

In practical theoretical calculations of the quarkonium production 
and decay rates, a number of significant uncertainties arise. In many
instances, the series in $\alpha_s$ and in $v$ of Eq.~(\ref{prod-fact}) 
converge slowly, and the uncertainties from their
truncation are large---sometimes of order 100\%. In addition, the matrix
elements are often poorly determined, either from phenomenology or
lattice measurements, and the important linear combinations of matrix
elements vary from process to process, making tests of universality
difficult. There are also large
uncertainties in the heavy-quark masses (approximately 10\% for $m_c$
and 5\% for $m_b$, for the mass ranges used in the calculations) that 
can be very significant for
quarkonium rates proportional to a large power of the mass.

\subsection{Experimental Tests of NRQCD Factorization}

Here, we give a brief review of some of the successes of NRQCD, as well
as some of the open questions.  We concentrate on hadroproduction
results for both unpolarized and polarized production.  We also discuss
briefly some recent two-photon, $e^+ e^-$, and photoproduction results.

\begin{figure}
\begin{center}
\epsfig{figure=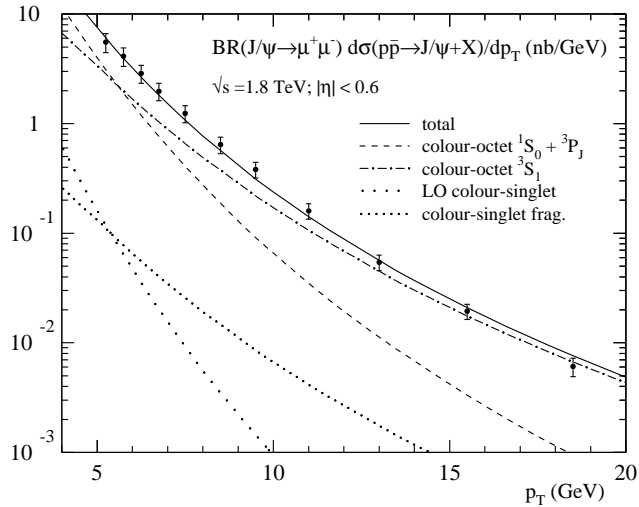,width=10cm}
\caption{$J/\psi$ cross section as a function of $p_T$. The data points 
are from the CDF measurement \cite{Abe:1997jz}. The solid curve is 
the NRQCD-factorization fit to the data given in Ref.~\cite{Kramer:2001hh}. 
The other curves give various contributions to the NRQCD-factorization 
fit. From Ref.~\cite{Kramer:2001hh}.}
\label{fig-tevatron}
\end{center}
\end{figure}

Using the NRQCD-factorization approach, one can obtain a good fit to the
high-$p_T$ CDF data \cite{Abe:1997jz}, while the color-singlet model
under predicts the data by more than an order of magnitude. (See
Fig.~\ref{fig-tevatron}.)  The $p_T$ dependence of the unpolarized
Tevatron charmonium data has been studied under a number of model
assumptions, including LO collinear factorization, parton-shower
radiation, $k_T$ smearing, and $k_T$ factorization. (See
Ref.~\cite{Kramer:2001hh} for a review.)

Several uncertainties in the theoretical predictions affect the
extraction of the NRQCD char\-mo\-ni\-um-production matrix elements from
the data. There are large uncertainties in the theoretical predictions
that arise from the choices of the factorization scale, the
renormalization scale, and the parton distributions. The  extracted
values of the octet matrix elements are very sensitive to the
small-$p_T$ behavior of the cross section and this, in turn, leads to a
sensitivity to the behavior of the small-$x$ gluon distribution.
Furthermore, the effects of multiple soft-gluon emission are important,
and their omission in the fixed-order perturbative calculations leads to
overestimates of the matrix elements. Effects of higher-order
corrections in $\alpha_s$ are a further uncertainty in the theoretical
predictions. Similar theoretical uncertainties arise in the extraction
of the NRQCD production matrix elements for the $\Upsilon$
\cite{Braaten:2000cm} states, but, owing to large statistical
uncertainties, they are less significant for the fits than in the
charmonium case.

\begin{figure}
\begin{tabular}{cc}
\includegraphics[width=8cm]{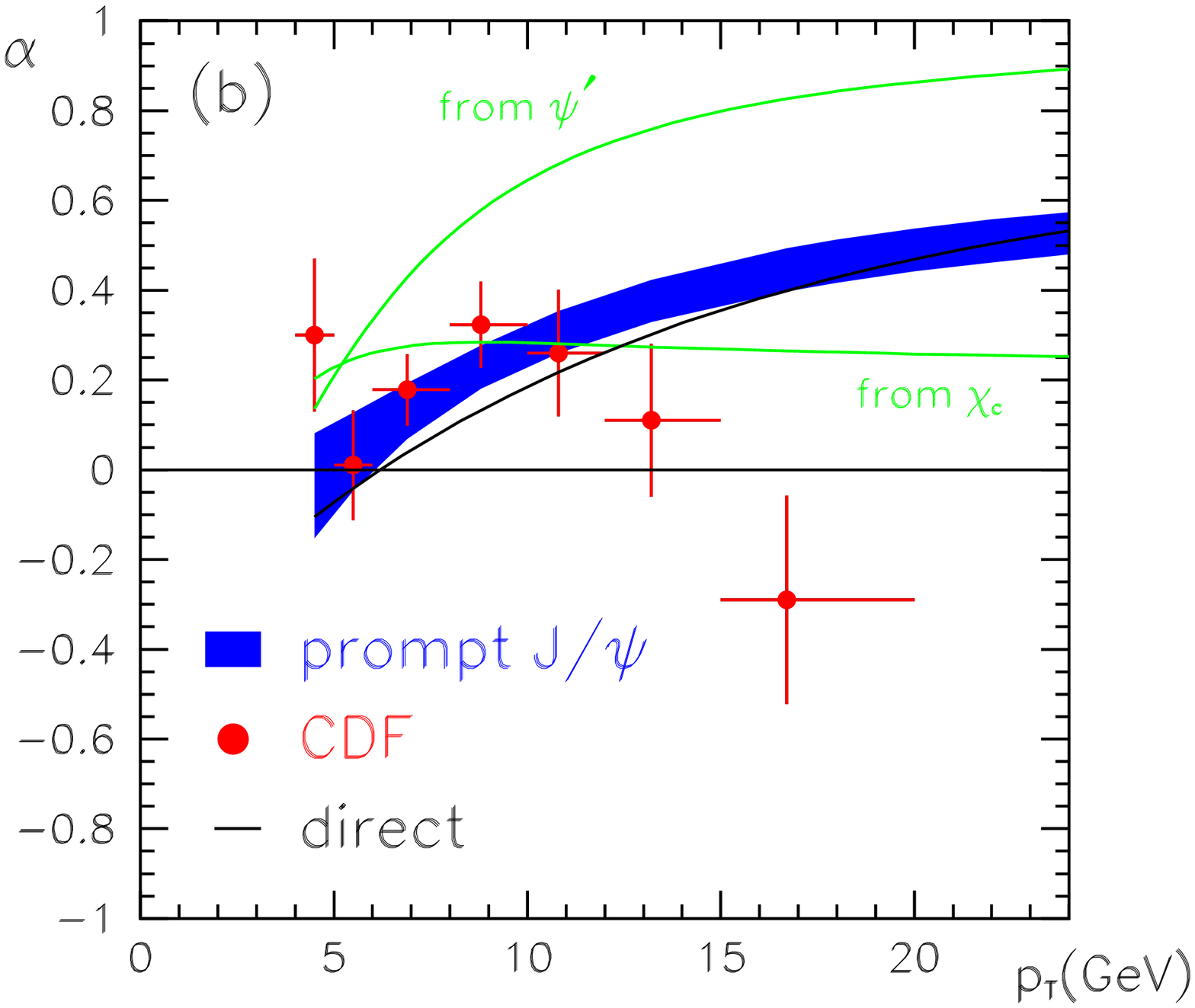}&
\includegraphics[width=8cm]{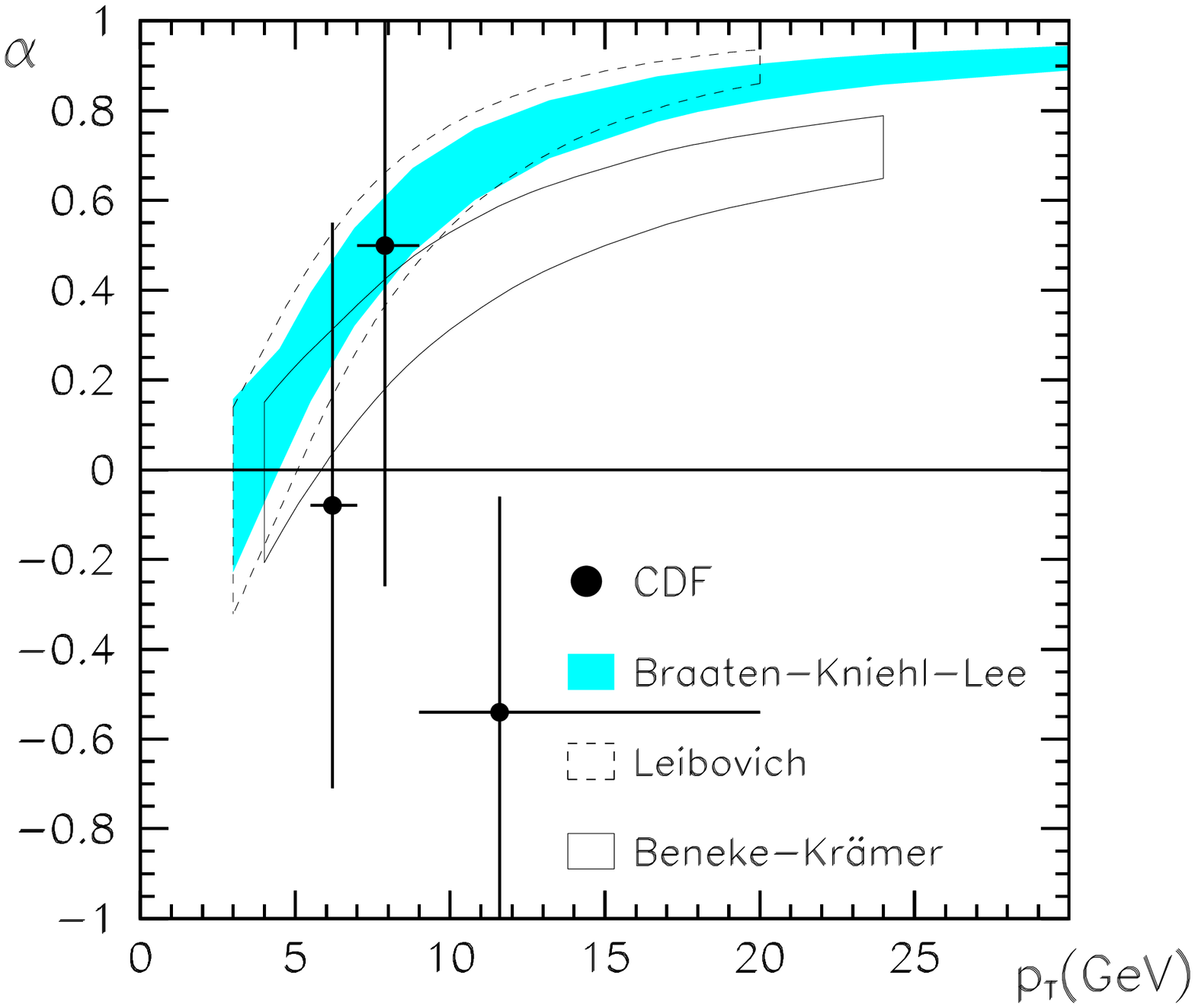}
\end{tabular}
\caption{Left-hand side: $J/\psi$ polarization at the Tevatron. 
The band is the total NRQCD-factorization prediction. The other curves 
give the contributions from feeddown from higher charmonium states. 
Right-hand side: $\psi'$ polarization at the Tevatron. 
The bands give various NRQCD-factorization predictions. The data 
points are from the CDF measurement \cite{Affolder:2000nn}. From 
Ref.~\cite{Braaten:1999qk}.}
\label{fig-pol}
\end{figure}

At large $p_T$ ($p_T\gsim 4m_c$ for the $J/\psi$) the dominant
quarkonium-production mechanism is gluon fragmentation into a
$Q\overline Q$ pair in a ${}^3S_1$ color-octet state.
The fragmenting gluon is nearly on mass shell
and is, therefore, transversely polarized. Furthermore, the velocity-scaling
rules predict that the color-octet $Q\overline Q$ state retains its
transverse polarization as it evolves into $S$-wave quarkonium
\cite{Cho:1994ih}, up to corrections of relative order $v^2$. Radiative
corrections, color-singlet production, and feeddown from higher states
can dilute the quarkonium polarization
\cite{Beneke:1995yb,Leibovich:1996pa,Beneke:1996yw,Braaten:1999qk,%
Kniehl:2000nn}. Despite this
dilution, a substantial polarization is expected at large $p_T$.  Its
detection would be a ``smoking gun'' for the presence of color-octet
production. In contrast, the color-evaporation model predicts no
quarkonium polarization. The CDF measurement of the $J/\psi$ and $\psi'$
polarization as a function of $p_T$ \cite{Affolder:2000nn} is shown in
Fig.~\ref{fig-pol}, along with the NRQCD factorization prediction
\cite{Leibovich:1996pa,Beneke:1996yw,Braaten:1999qk}. The analysis of 
$\psi'$ polarization is simpler
than for the $J/\psi$, since feeddown does not play a r\^ole. However, the
statistics are not as good for the $\psi'$.
The degree of polarization is $\alpha=(1-3\xi)/(1+\xi)$, where $\xi$ is 
the fraction of events with longitudinal polarization.
$\alpha=1$ corresponds to 100\% transverse polarization, and
$\alpha=-1$ corresponds to 100\% longitudinal polarization. The observed
polarization is in relatively good agreement with the prediction, except
in the highest $p_T$ bin, although the prediction of increasing
polarization with increasing $p_T$ is not in evidence.

Because the polarization depends on a ratio of matrix elements, some of
the theoretical uncertainties are reduced compared with those in the
production cross section, and, so, the polarization is probably not
strongly affected by multiple soft-gluon emission or $K$ factors.
Contributions of higher order in $\alpha_s$ could conceivably change the
rates for the various spin states by a factor of two. Therefore, it is
important to carry out the NLO calculation, which involves significant
computational difficulties. It is known that order-$v^2$ corrections to
parton fragmentation into quarkonium can be quite large
\cite{bodwin-lee}. If spin-flip corrections to the NRQCD matrix
elements, which are nominally suppressed by powers of $v$, are also
large, perhaps because the velocity-scaling rules need to be modified,
then spin-flip contributions could significantly dilute the $J/\psi$
polarization. Nevertheless, in the context of NRQCD, it is difficult to
see how there could not be substantial charmonium polarization for
$p_T>4m_c$.

Compared to the $J/\psi$-polarization prediction, the
$\Upsilon$-polarization prediction has smaller $v$-ex\-pan\-sion
uncertainties. However, because of the higher $\Upsilon$ mass, it is
necessary to go to higher $p_T$ to insure that fragmentation dominates
and that there is substantial polarization. Unfortunately, the current
Tevatron data run out of statistics in this high-$p_T$ region.
CDF finds that $\alpha=-0.12\pm 0.22$ for $8 <p_T<20\hbox{ GeV}$
\cite{Acosta:2001gv}, which is consistent with both
the NRQCD-factorization prediction \cite{Braaten:2000gw} and the
zero-polarization prediction of the CEM. There are also discrepancies
between the polarizations observed in fixed-target experiments and the
NRQCD predictions.

Calculations of inclusive $J/\psi$ and $\Upsilon$ production in $\gamma
\gamma$ collisions \cite{Klasen:2001cu,Cho:1995vv} have been compared
with LEP data
\cite{Todorova-Nova:2001pt,Chapkine:2002,Alexander:1995vh}. Both the
$J/\psi$ and $\Upsilon$ measurements favor the NRQCD predictions over
those of the color-singlet model.

Belle \cite{Abe:2001za} and BaBar \cite{Aubert:2001pd} have also
measured the $J/\psi$ total cross sections in $e^+e^- \rightarrow J/\psi
X$. The results of the two experiments are incompatible with each other,
but they both seem to favor NRQCD over the color-singlet model. A
surprising new result from Belle \cite{Abe:2002rb} is that most of the
produced $J/\psi$'s are accompanied by an additional $c\overline c$
pair: $\sigma(e^+e^-\rightarrow J/\psi\,c\overline c)
/\sigma(e^+e^-\rightarrow J/\psi\, X)=0.59^{+0.15}_{-0.13}\pm 0.12$.
Perturbative QCD plus the color-singlet model predict that this ratio
should be about $0.1$ \cite{Cho:1996cg}. There seems to be a major
discrepancy between theory and experiment. However, the
order-$\alpha_s^2$ calculation lacks color-octet contributions, including
those that produce $J/\psi\, c\overline c$. Although these contributions
are suppressed by $v^4 \approx 0.1$, it is possible that the
short-distance coefficients are large.  In other results, the angular
distributions favor NRQCD, but the polarization measurements show no
evidence of the transverse polarization that would be expected in
color-octet production. However, the center-of-mass momentum is rather
small, and, hence, one would not expect the polarization to be large.

Quarkonium production has also been measured in inelastic photoproduction
\cite{Merkel:qu,Bertolin:2000xz} and deep-inelastic scattering (DIS)
\cite{Meyer:1998fw,Mohrdieck:2000mk} at HERA. The NRQCD calculation
deviates from the data near large photon-momentum fractions, owing to
the large LO color-octet contribution. The NLO color-singlet result
agrees with the data over all momentum fractions, as well as with the
data as a function of $p_T$. See Ref.~\cite{Kramer:2001hh} for a more
complete review.  In the case of deep-inelastic scattering, the $Q^2$
and $p_T$ dependences are in agreement with NRQCD, but the results are
more ambiguous for the dependence on the longitudinal momentum fraction.

\subsection{Quarkonium Production in Nuclear Matter}

The existing factorization ``theorems'' for quarkonium production in
hadronic collisions are for cold hadronic matter. These theorems predict
that nuclear matter is ``transparent'' for $J/\psi$ production at large
$p_T$. That is, at large $p_T$, all of the nuclear effects are contained
in the nuclear parton distributions. The corrections to this
transparency are of order $(m_Q v/p_T)^2$ for unpolarized cross
sections and of order $m_Qv/p_T$ for polarized cross sections.

The effects of transverse-momentum kicks from multiple elastic
collisions between active partons and spectators in the nucleons are
among those effects that are suppressed by $(m_Q v/p_T)^2$. Nevertheless,
these multiple-scattering effects can be important because the
production cross section falls steeply with $p_T$ and because the number
of scatterings grows linearly with the path length through
nuclear matter. Such elastic interactions can be expressed in terms of
eikonal interactions \cite{Bodwin:1988fs} or higher-twist matrix
elements \cite{Qiu:2001hj}.

Inelastic scattering of quarkonium by nuclear matter is also an
effect of higher order in $(m_Q v/p_T)^2$. However, it can become
dominant when the amount of nuclear matter that is traversed by the
quarkonium is sufficiently large. Factorization breaks down when
\begin{equation}
L\gsim {{\rm min}(z_Q,z_{\overline Q})P_H^2\over 
M_A k_T^2({\rm tot})} \, \, ,
\end{equation}
where $L$ is the length of the quarkonium path in the nucleus, $M_A$ is
the mass of the nucleus, $z$ is the parton longitudinal momentum
fraction, $P_H$ is the momentum of the quarkonium in the
parton CM frame, and $k_T({\rm tot})$ is the accumulated transverse
momentum ``kick'' from passage through the nuclear matter. This
condition for the breakdown of factorization is similar to
``target-length condition'' in Drell-Yan production
\cite{Bodwin:1981fv,Bodwin:1984hc}. Such a breakdown 
is observed in the Cronin effect at low $p_T$ and in
Drell-Yan production at low $Q^2$, where the cross section is
proportional to $A^\alpha$, and $\alpha < 1$.

It is possible that multiple-scattering effects may be
larger for color-octet production than for color-singlet production.
In the case of color-octet production, the pre-quarkonium $Q\overline Q$
system carries a nonzero color charge and, therefore, has a larger 
amplitude to exchange soft gluons with spectator partons.

At present, there is no complete, rigorous theory to account for all of
the effects of multiple scattering and we must resort to
``QCD-inspired'' models. A reasonable requirement for models is that
they be constructed so that they are compatible with the factorization
result in the large-$p_T$ limit. Many models treat interactions of the
pre-quarkonium with the nucleus as on-shell (Glauber) scattering. 
This assumption should be examined carefully, as on-shell
scattering is known, from the factorization proofs, not to be a valid
approximation in leading order in $(m_Q v/p_T)^2$.

\section{NRQCD Predictions for the LHC
}
\label{quarkon.nrqcd2}

In this section, we shall use the formalism of NRQCD to give predictions
for quarkonium production in the LHC energy range. We rewrite
the cross section in Eq.~(\ref{prod-fact}) for the inclusive production of a 
charmonium state $H$ as follows:
\begin{equation} 
\sigma(H) \;=\; \sum_n
\sigma^{(Q\overline Q)_n} \;
        \langle {\cal O}^{H}_n \rangle \, \, ,
\label{sig-fact}
\end{equation}
where $\sigma^{(Q \overline Q)_n} = F_n(\Lambda)/m_Q^{d_n-4}$,
$\langle {\cal O}_n^H \rangle = \langle 0| {\cal O}_n^H|0\rangle$, 
and $n$ runs over
all the color and angular momentum states of the $Q \overline Q$ pair.
The cross sections $\sigma^{(Q \overline Q)_n}$
can be calculated in perturbative QCD.
All dependence on the final state $H$ is contained in the 
nonperturbative NRQCD matrix elements 
$\langle {\cal O}^{H}_n \rangle$.

The most important matrix elements for
$J/\psi=\psi(1S)$ and $\psi'=\psi(2S)$ production
can be reduced to the color-singlet parameter 
$\langle {\cal O}^{\psi(nS)}_1(^3S_1) \rangle$ 
and the three color-octet parameters
$\langle {\cal O}^{\psi(nS)}_8(^3S_1) \rangle$,
$\langle {\cal O}^{\psi(nS)}_8(^1S_0) \rangle$, and
$\langle {\cal O}^{\psi(nS)}_8(^3P_0) \rangle$.
Two of the three color-octet matrix elements only appear in the linear 
combination
\begin{eqnarray}
M_k^{\psi(nS)} = (k/m_c^2)\langle {\cal O}^{\psi(nS)}_8(^3P_0) \rangle 
+ \langle
{\cal O}^{\psi(nS)}_8(^1S_0)\rangle \, \, . \label{mk}
\end{eqnarray}  
The value of $k$ is sensitive to the $p_T$
dependence of the fit.  At the Tevatron, $k \approx 3$.  Fits to fixed-target
total cross sections give larger values, $k \approx \hbox{6--7}$ 
\cite{Beneke:1996tk}.
The most important matrix elements for $\chi_{cJ}$ production
can be reduced to a color-singlet parameter
$\langle {\cal O}^{\chi_{c0}}_1(^3P_0) \rangle$
and a single color-octet parameter
$\langle {\cal O}^{\chi_{c0}}_8(^3S_1) \rangle$.
These matrix elements are sufficient to
calculate the prompt $J/\psi$ cross section to leading order in 
$\alpha_s$ and to order $v^4$ relative to 
the color-singlet contribution.  
\begin{table}
\begin{center}
\caption{Matrix elements for charmonium production. Note that here
$\langle {\cal O}_1^H \rangle = \langle {\cal
O}_1^{\psi(nS)}(^3S_1) \rangle$ for $J/\psi$ and $\psi'$, but $\langle
{\cal O}_1^H \rangle=\langle {\cal O}_1^{\chi_{c0}}(^3P_0) \rangle$ for
$\chi_{c0}$.  Uncertainties are statistical only. From
Ref.~\cite{Braaten:1999qk}.}
\label{tab:me-1}
\renewcommand{\arraystretch}{1.5}
$$
\begin{array}{|c|cccc|}
\hline\hline
 H & \langle {\cal{O}}_1^{H} \rangle  & \langle
 {\cal{O}}_8^{H}(^3S_1) \rangle  & k &
 M_k^{H} \\ \hline
J/\psi & 1.3 \pm 0.1~{\rm GeV^3} & (4.4 \pm 0.7)\times 10^{-3}~{\rm GeV}^3 &  
3.4 & (8.7 \pm 0.9)\times 10^{-2}~{\rm GeV}^3 \\[-1mm] 
\psi' & 0.65 \pm 0.06~{\rm GeV^3} & (4.2 \pm 1.0)\times 10^{-3}~{\rm GeV}^3 & 
3.5 & (1.3 \pm 0.5)\times 10^{-2}~{\rm GeV}^3 
 \\[-1mm]
\chi_{c0} & (8.9 \pm 1.3)\times 10^{-2}~{\rm GeV^5} & 
(2.3 \pm 0.3)\times 10^{-3}~{\rm GeV}^3 & & \\[1mm] \hline \hline
\end{array}
$$
\renewcommand{\arraystretch}{1.0}
\end{center}
\end{table}

In $pp$ collisions, different partonic processes for 
$Q \overline Q$ production dominate in different $p_T$ ranges.
If $p_T$ is of order $m_Q$, fusion processes dominate, and, so, 
the $Q \overline Q$ pair is produced in the hard-scattering process.
These contributions can be written in the form
\begin{equation}
\sigma_{\rm Fu}(H) = \sum_{i,j} \int dx_1 \, dx_2 \, 
f_{i/A}(x_1,\mu^2) f_{j/B}(x_2,\mu^2)
\hat \sigma_{ij}^{(Q \overline Q)_n} \;
\langle {\cal O}^{H}_n \rangle \, \,  ,
\label{sig-fusion}
\end{equation}
where $A$ and $B$ are the incoming hadrons or nuclei.  
In Eq.~(\ref{sig-fusion}),
we include the parton processes $i j \rightarrow Q \overline Q \, X$,
where $ij=gg,q \overline q, qg$ and $\overline q g$, and $q=u,d,s$.
The relevant partonic cross sections $\hat \sigma_{ij}^{(Q \overline Q)_n}$
are given in Refs.~\cite{Cho:1995vh,Cho:1995ce}.

For $p_T\gg m_Q$, the dominant partonic process is gluon 
fragmentation through the color-octet ${}^3S_1$ channel.
This contribution can be expressed as
\begin{equation}
\sigma_{\rm Fr}(H) = \sum_{i,j}\int dx_1 \, dx_2 \, dz
f_{i/A} (x_1,\mu^2) f_{j/B} (x_2,\mu^2)
\hat \sigma_{ij}^{g}
D_g^{(Q \overline Q)_8({}^3S_1)} (z,\mu_{\rm Fr}^2) \; 
\langle {\cal O}^{H}_n \rangle \,
\, ,
\label{sig-frag}
\end{equation}
where $D_g^{(Q \overline Q)_n({}^3S_1)}(z,\mu_{\rm Fr}^2)$ is the
fragmentation function for a gluon fragmenting into a $Q\overline Q$ pair,
$P/z$ is the momentum of the fragmenting gluon, $P$ is the momentum of
the $Q\overline Q$ pair, and $\mu_{\rm Fr}$ is the fragmentation scale. The
fragmentation process scales as $d \hat \sigma/dp_T^2 \sim
1/p_T^4$~\cite{Braaten:1993rw,Braaten:1994vv}. The fragmentation process
is actually included in the fusion processes of Eq.~(\ref{sig-fusion}).
In the limit $p_T\gg m_Q$, the fusion processes that proceed through
$g^*\to (Q\overline Q)_8({}^3S_1)$ are well-approximated by the expression
(\ref{sig-frag}).  At large $p_T$, one can evolve the fragmentation
function in the scale $\mu_{\rm Fr}$, thereby resumming large logarithms
of $p_T^2/m_Q^2$. Such a procedure leads to a smaller short-distance
factor \cite{Beneke:1996yw} and a more accurate prediction at large
$p_T$ than would be obtained by using the fusion cross section
(\ref{sig-fusion}). However, in our calculations, we employ the fusion
cross section (\ref{sig-fusion}), which leads to systematic
over-estimation of the cross section at large $p_T$.

In order to predict the cross section for prompt $J/\psi$ production
(including $\chi_c$ and $\psi'$ feeddown) at the LHC, we need the values
of the NRQCD matrix elements.  There have been several previous
extractions of the color-octet matrix elements
\cite{Beneke:1996yw,Braaten:1999qk,%
Cho:1995vh,Cho:1995ce,Cacciari:1995yt,Kniehl:1998qy,Kniehl:1999vf} from
the CDF $J/\psi$, $\chi_c$ and $\psi'$ $p_T$ distributions
\cite{Abe:1997jz,Abe:1997yz}. We use the matrix elements given in
Ref.~\cite{Braaten:1999qk}, which are shown in
Table~\ref{tab:me-1}. Our calculations are based on the MRST LO
parton distributions \cite{Martin:1998np}. 
\begin{figure}
\begin{center}
\includegraphics[width=8cm]{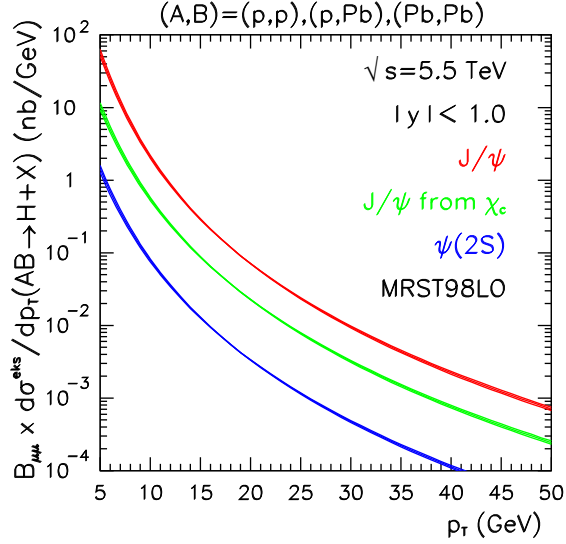}
\end{center}
\caption{\label{lee.fig1} Differential cross sections per nucleon
multiplied by leptonic branching fractions
for prompt $J/\psi$ (upper curves),
$J/\psi$ from $\chi_c$ (middle curves), and
prompt $\psi(2S)$ (lower curves) in $pp$, $p$Pb, and Pb+Pb collisions
at $\sqrt{s}=5.5$~TeV. 
The EKS98 parametrization~\cite{Eskola:1998iy,Eskola:1998df} 
is employed for $p$Pb and Pb+Pb collisions.
}
\end{figure}
In calculating the cross section per nucleon
for prompt $J/\psi$ production in $pA$ or $AA$ collisions, we take
$f_{i/A}=f_{i/p} R^A_i$.  We employ the EKS98 
parametrization~\cite{Eskola:1998iy,Eskola:1998df}
for the nuclear shadowing ratio $R^A_i$.
We evolve $\alpha_s$ at one-loop accuracy, and 
we set $\mu = (4 m_c^2 + p_T^2)^{1/2}$ and $m_c=1.5$ GeV.

There are several sources of uncertainty in our predictions for the cross
sections. There are large uncertainties in the NRQCD matrix elements
themselves. The errors shown in Table \ref{tab:me-1} are 
statistical only. There are additional large uncertainties in the matrix 
elements that arise from truncations of the series in $\alpha_s$ and 
$v$ in the theoretical expressions that
are used to extract the matrix elements. The matrix elements $\langle
{\cal O}_8(^1S_0) \rangle$ and $\langle {\cal O}_8(^3P_0) \rangle$ are
fixed by the data only in the linear combination $M_k^H$. In the present
calculation, we take $\langle {\cal O}_8(^1S_0) \rangle=xM_k^H$ and
$\langle {\cal O}_8(^3P_0) \rangle/m_c^2=(1-x) M_k^H/k$, use the
values of $k$ given in Table \ref{tab:me-1}, and choose $x=1/2$.
Variation of $x$ between $0$ and $1$ affects the cross sections at 
low $p_T$ by amounts on the order of 5\%. There are
additional uncertainties in the predicted cross sections that arise from
the choices of the parton distributions, the charm-quark mass $m_c$, and
the scale $\mu$. Because they affect the matrix-element fits,
these uncertainties are highly correlated with those of the matrix 
elements. We have not tried to estimate their effects on the
predicted cross sections.
        
\begin{figure}
\begin{tabular}{cc}
\includegraphics[width=8cm]{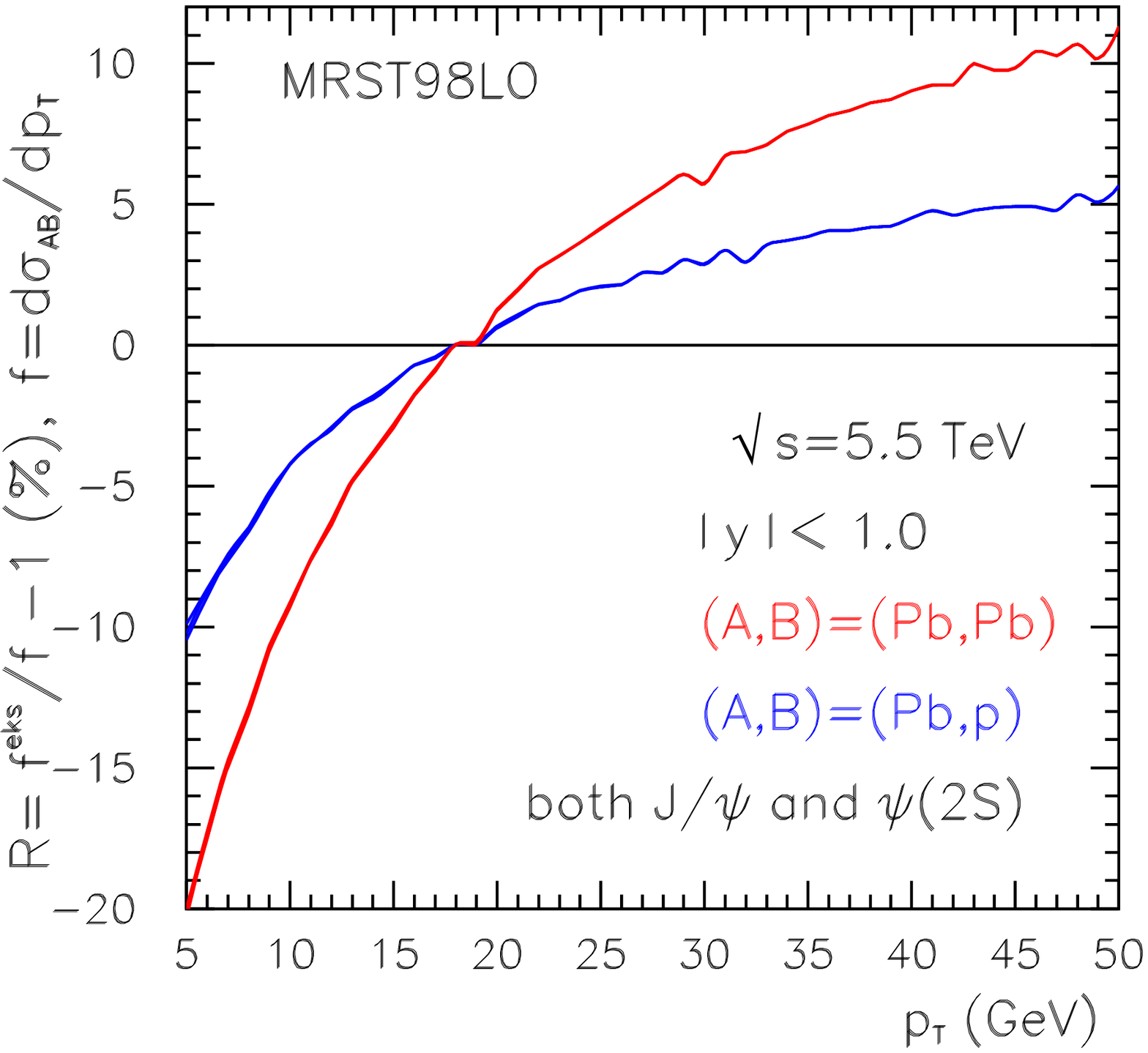}&
\includegraphics[width=8cm]{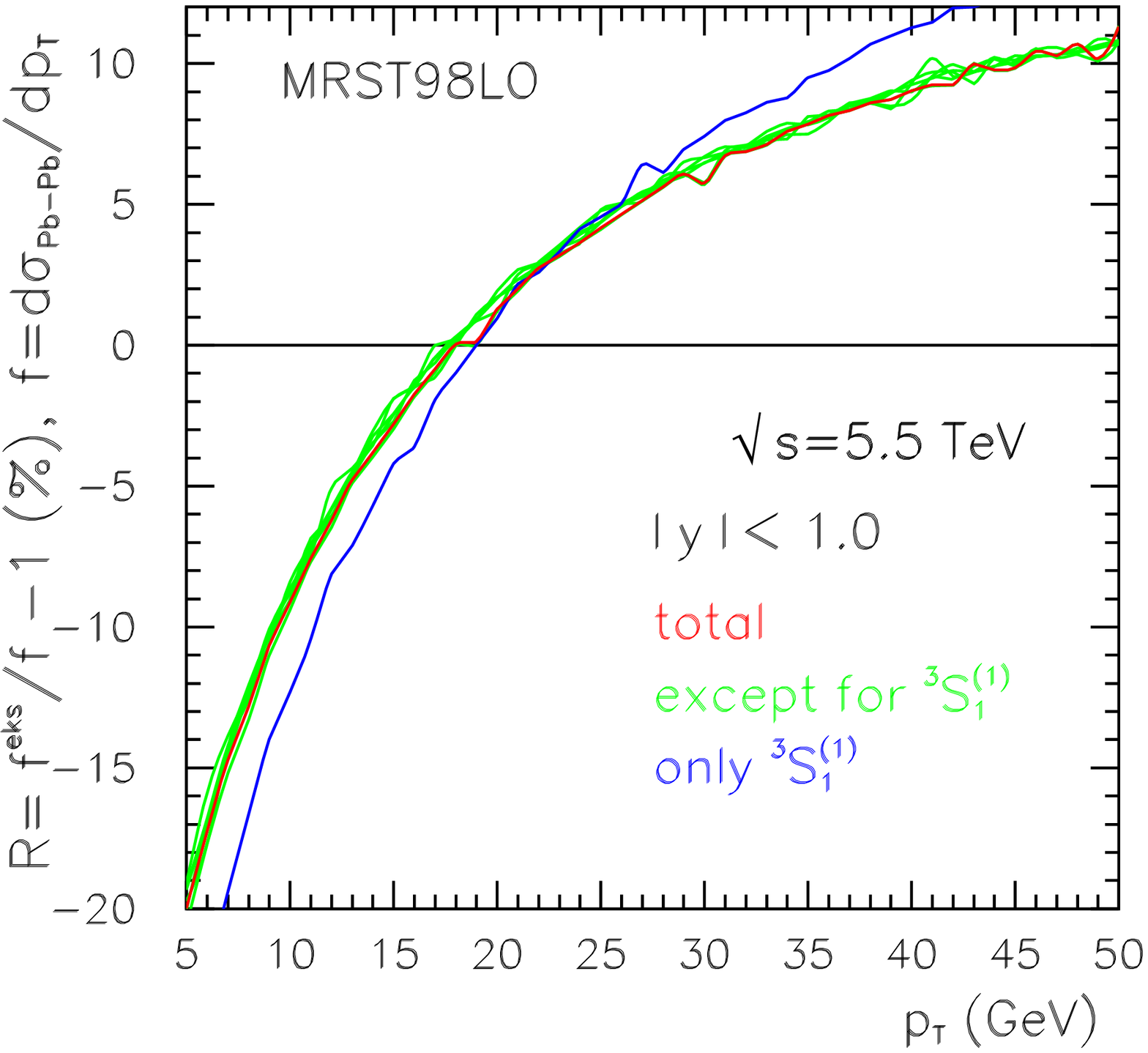}
\end{tabular}
\caption{\label{lee.fig2} 
The $p_T$ dependence of $R_{AB}$ [Eq.~(\ref{R})].
(a) We compare the results for $p$Pb and Pb+Pb collisions.  (The $p_T$ 
dependence is stronger in the Pb+Pb result.)
(b) We show the dependence of $R_{AB}$ on the various production channels
in Pb+Pb collisions at $\sqrt{s}=5.5$~TeV.
}
\end{figure}

In Fig.~\ref{lee.fig1}, we show the $p_T$ distributions per nucleon
multiplied by the dilepton branching fractions for prompt $J/\psi$
(upper curves), $J/\psi$ from $\chi_c$ decays (middle curves), and
prompt $\psi'$ (lower curves) at $\sqrt{s}=5.5$~TeV. For $p$Pb and Pb+Pb
collisions, we use the EKS98
parametrization~\cite{Eskola:1998iy,Eskola:1998df} to account for the
effect of nuclear shadowing. The $pp$, $p$Pb, and Pb+Pb results
essentially lie on top of each other in Fig.~\ref{lee.fig1}, owing to the
many decades covered in the plot.

In order to display small differences between the distributions, we
define the function $R_{AB}$:
\begin{eqnarray}
R_{AB}(p_T)=\frac{d\sigma_{AB}/dp_T - d\sigma_{pp}/dp_T}
                 {d\sigma_{pp}/dp_T} \, \, .
\label{R}
\end{eqnarray}
In Fig.~\ref{lee.fig2}, we present $R_{AB}$ as a function of $p_T$. As
is shown in Fig.~\ref{lee.fig2}(a), nuclear shadowing increases the
cross section at large $p_T$ and decreases it at small $p_T$. The
deviation of the Pb+Pb cross section from the $pp$ cross section is
twice as large as that seen in the case of $p$Pb collisions. In order to
investigate the dependence of the shadowing effect on the short-distance
cross sections that arise in hadroproduction of $S$-wave charmonium
states in Pb+Pb collisions, we plot $R_{AB}$ for all channels separately.
[See Fig.~\ref{lee.fig2}(b).] Even though the $p_T$ dependence the
contribution to the cross section of the color-octet $^3S_1$ channel is
quite different from those of the color-octet $^1S_0$ and $^3P_J$
channels, all three channels show the same nuclear effect. The only
channel that shows a slightly different behavior is the color-singlet
channel, which gives a negligible contribution to the cross section.
While the differential cross sections in Fig.~\ref{lee.fig1} are
strongly dependent on the nonperturbative NRQCD matrix elements,
$R_{AB}$ is almost independent of the matrix elements, making it a good
observable for studying nuclear shadowing at the LHC.

The $\Upsilon$ rates are somewhat more difficult to calculate because of
the many feeddown contributions.  The matrix elements are also not
particularly well known. Since it is unlikely that all the different
contributions can be disentangled, we follow the approach of
Ref.~\cite{Braaten:2000cm} and compute the inclusive $\Upsilon(nS)$
production cross section
\begin{eqnarray}
d\sigma(\Upsilon(nS))_{\rm inc} & = & 
d \sigma^{(b \overline b)_1(^3S_1)} 
        \langle {\cal O}_1^{\Upsilon(nS)}(^3S_1) \rangle_{\rm inc}
+ \sum_J d \sigma^{(b \overline b)_1(^3P_J)} 
        \langle {\cal O}_1^{\Upsilon(nS)}(^3P_J) \rangle_{\rm inc}
\nonumber
\\ 
&& + \mbox{} d \sigma^{(b \overline b)_8(^3S_1)}
        \langle {\cal O}_8^{\Upsilon(nS)}(^3S_1) \rangle_{\rm inc}
+ d \sigma^{(b \overline b)_8(^1S_0)}
        \langle {\cal O}_8^{\Upsilon(nS)}(^1S_0) \rangle_{\rm inc}
\nonumber
\\
&& 
+ \left( \sum_J (2J+1) d \sigma^{(b \overline b)_8(^3P_J)} \right)
        \langle {\cal O}_8^{\Upsilon(nS)}(^3P_0) \rangle_{\rm inc} \, \, ,
\label{sig-total}
\end{eqnarray}
where the last term makes use of heavy-quark spin symmetry to relate all
of the octet ${}^3P_J$ matrix elements to the octet ${}^3P_0$ matrix element.
The ``inclusive'' matrix elements are defined by
\begin{equation}
\langle {\cal O}_i^{\Upsilon(nS)}(n) \rangle_{\rm inc} =
\sum_H B_{H \to \Upsilon(nS)} \langle {\cal O}_i^H(n) \rangle \, \, ,
\label{O-total}
\end{equation}
where $i=1$ or 8 for singlet or octet, respectively.
\begin{table}
\begin{center}
\caption{Inclusive color-singlet matrix elements for bottomonium
production.  The errors on the $^3S_1$ matrix elements come from
estimates of the $\Upsilon(nS)$ decay rate to lepton pairs.  The errors
on the $^3P_J$ states come from an average over potential-model
estimates.  The inclusive matrix elements are a linear combination of
branching ratios, as in Eq.~(\protect\ref{O-total}).  The $S$-state
matrix elements are in units of GeV$^3$ while the $P_J$-state matrix
elements are in units of GeV$^5$.  From Ref.~\cite{Braaten:2000cm}.}
\label{upssing}
\renewcommand{\arraystretch}{1.5}
$$
\begin{array}{|c|cccc|}
\hline\hline
 H & \langle {\cal{O}}_1^{H}(^3S_1) \rangle_{\rm inc} & \langle
 {\cal{O}}_1^H(^3P_0) \rangle_{\rm inc}  & \frac{1}{3} \langle 
{\cal{O}}_1^H(^3P_1) \rangle_{\rm inc} & \frac{1}{5} \langle 
{\cal{O}}_1^H(^3P_2) \rangle_{\rm inc}  \\ \hline
\Upsilon(3S) & 4.3 \pm 0.9 & 0 & 0 & 0 \\[-1mm] 
\Upsilon(2S) & 5.0 \pm 0.7 & 0.12 \pm 0.06 & 0.55 \pm 0.15 & 0.42 \pm 0.10
 \\[-1mm]
\Upsilon(1S) & 12.8 \pm 1.6 & < 0.2 & 1.23 \pm 0.25 & 0.84 \pm 0.15
 \\[-1mm] \hline \hline
\end{array}
$$
\renewcommand{\arraystretch}{1.0}
\end{center}
\end{table}
The sum over $H$ includes the $\Upsilon(nS)$ as well as all higher states
that can decay to $\Upsilon(nS)$.  The branching ratio for $H \to H'$
decays is $B_{H \to H'}$ with $B_{H \to H} \equiv 1$.  Only $\chi_b(1P)$
and $\chi_b(2P)$ decays are included; the possibility of feeddown from
the as-yet unobserved $\chi_b(3P)$ states is neglected.  In the linear
combination $M_k^{\Upsilon(nS)}$, the color-octet matrix element from
the $^3P_0$ state is neglected, and, so, $M_k^{\Upsilon(nS)} = \langle
{\cal O}_8^{\Upsilon(nS)}(^1S_0) \rangle_{\rm inc}$.  We use $m_b =
4.77$ GeV and the MRST LO parton distributions. The values of the
inclusive color-singlet matrix elements are given in
Table~\ref{upssing}, and the values of the inclusive color-octet matrix
elements, from Ref.~\cite{Braaten:2000cm}, are given in
Table~\ref{upsoct}.

\begin{table}
\begin{center}
\caption{Inclusive color-octet matrix elements for bottomonium
production. The matrix elements were fit using the MRSTLO parton 
distributions. The first set of error bars is from $\chi^2$ fits to the
$\Upsilon$ $p_T$ distributions in the region $p_T>8$ GeV.  The second
set is associated with the variation of the scales and corresponds to
multiplying $\mu = \sqrt{m_b^2 + p_T^2}$ by 2 (upper error) and 0.5
(lower error).  The matrix elements are in units of $10^{-2}$ GeV$^3$.
From Ref.~\cite{Braaten:2000cm}.}
\label{upsoct}
\renewcommand{\arraystretch}{1.5}
$$
\begin{array}{|c|ccc|}
\hline\hline
 H & \langle {\cal{O}}_8^{H}(^3S_1) \rangle_{\rm inc} & \langle
 {\cal{O}}_8^H(^1S_0) \rangle_{\rm inc}  & \frac{5}{m_b^2} \langle
{\cal{O}}_1^H(^3P_0) \rangle_{\rm inc}  \\ \hline
\Upsilon(3S) & 3.7 \pm 1.7 ^{+1.7}_{-1.3} & 7.5 \pm 4.9 ^{+3.4}_{-2.5} & 0
 \\[-1mm] 
\Upsilon(2S) & 19.6 \pm 6.3 ^{+8.9}_{-6.5} & -8.7 \pm 11.1 ^{-2.4}_{+1.8} & 0
 \\[-1mm]
\Upsilon(1S) & 11.7 \pm 3.0 ^{+5.7}_{-4.2} & 18.1 \pm 7.2 ^{+11.4}_{-8.1} & 0
 \\[-1mm] \hline \hline
\end{array}
$$
\renewcommand{\arraystretch}{1.0}
\end{center}
\end{table}

\begin{figure}
\begin{center}
\includegraphics[width=8cm]{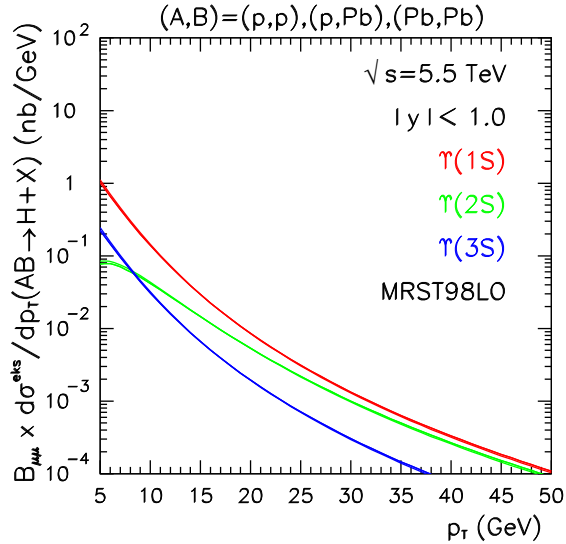}
\end{center}
\caption{\label{lee.fig3} Differential cross sections per nucleon
multiplied by leptonic branching fractions
for inclusive $\Upsilon(1S)$ (upper curves),
$\Upsilon(2S)$ (middle curves), and
prompt $\Upsilon(3S)$ (lower curves) in $pp$, $p$Pb, and Pb+Pb collisions
at $\sqrt{s}=5.5$~TeV. 
The EKS98 parametrization~\cite{Eskola:1998iy,Eskola:1998df} 
is employed for $p$Pb and Pb+Pb collisions.
}
\end{figure}

In Fig.~\ref{lee.fig3}, we show the $p_T$ distributions per nucleon
multiplied by the dilepton branching fractions for the 3 $\Upsilon$ $S$
states at $\sqrt{s}=5.5$~TeV.  The feeddown contributions are included
as in Eq.~(\ref{sig-total}).  For $p$Pb and Pb+Pb collisions, we use the
EKS98 parametrization~\cite{Eskola:1998iy,Eskola:1998df} in order to
account for the effects of nuclear shadowing. The $pp$, $p$Pb, and Pb+Pb
results lie essentially on top of each other in Fig.~\ref{lee.fig3}.

The unusual relative behavior of the $\Upsilon(2S)$ and $\Upsilon(3S)$
states at both low and high $p_T$ is due to the fact that the
bottomonium matrix elements are not very well determined.  For $p_T <
10$ GeV, the $\Upsilon(2S)$ cross section drops below the $\Upsilon(3S)$
cross section because the $\Upsilon(2S)$ has a large negative
color-octet matrix element. (See Table~\ref{upsoct}.) The short-distance
coefficients multiplying $M_k^{\Upsilon(nS)}$ are significant at low
$p_T$. Thus, there is a large cancellation between the octet $^3S_1$
matrix element and $M_k^{\Upsilon(nS)}$, which reduces the
$\Upsilon(2S)$ cross section in this region, causing it to drop below
the $\Upsilon(3S)$ cross section at low $p_T$. At the high-$p_T$ end of
the spectrum, the large value of the $^3S_1$ $\Upsilon(2S)$ color-octet
matrix element (Table~\ref{upsoct}) causes the $\Upsilon(2S)$ cross
section to approach that of the $\Upsilon(1S)$.  In this region, the
color-octet $^3S_1$ contribution dominates the other channels. Its large
matrix element gives the $\Upsilon(2S)$ an unreasonably large cross
section relative to that of the $\Upsilon(1S)$.  The $\Upsilon(2S)$ rate
at $p_T \approx 15$ GeV is more reasonable because the large and positive
$^3S_1$ contribution, and the large and negative $M_k^{\Upsilon(2S)}$
contribution nearly cancel each other.

\begin{figure}
\begin{tabular}{cc}
\includegraphics[width=8cm]{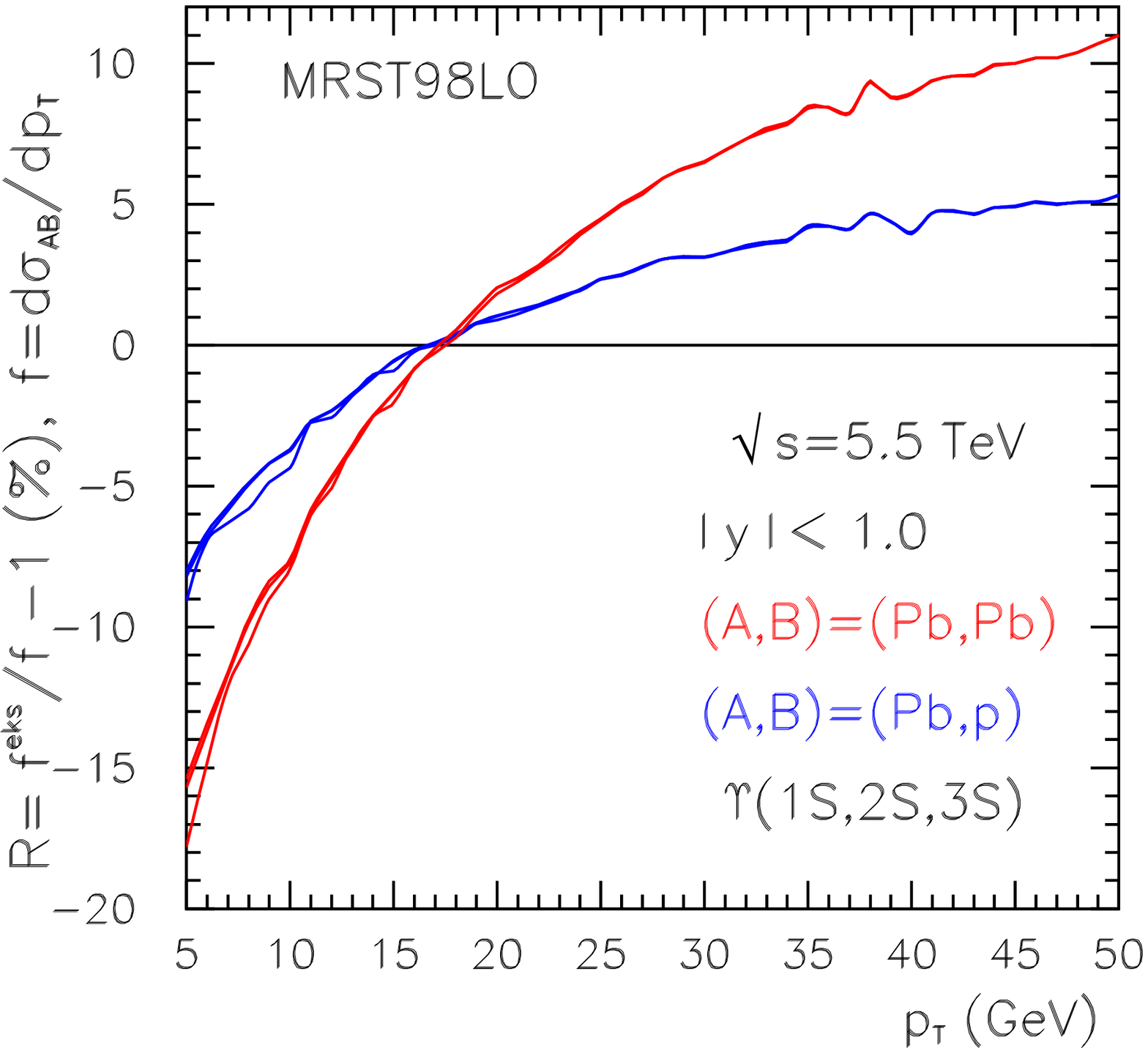}&
\includegraphics[width=8cm]{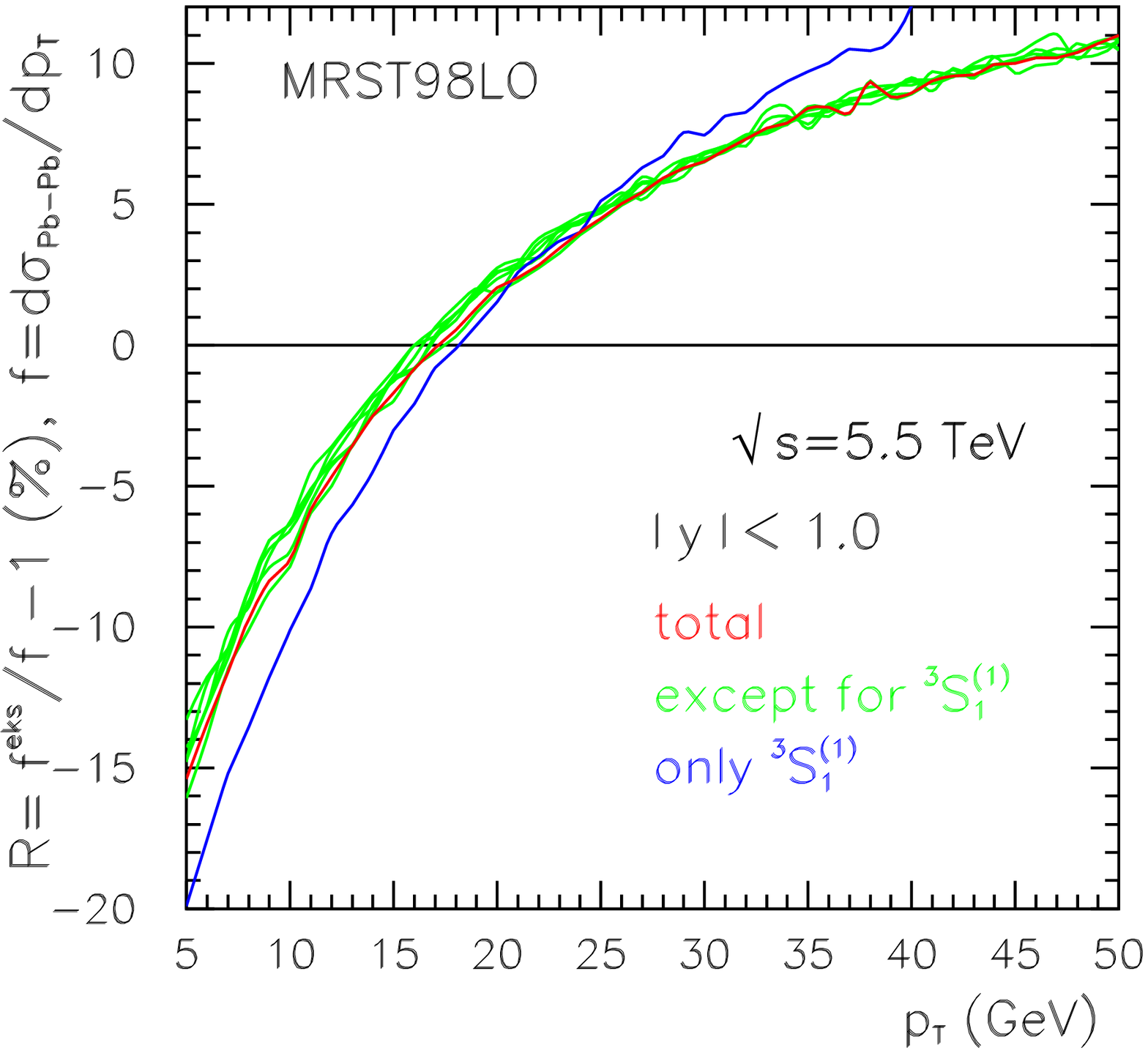}
\end{tabular}
\caption{\label{lee.fig4} 
The $p_T$ dependence of $R_{AB}$ [Eq.~(\ref{R})] for $\Upsilon$ 
production.
(a) We compare the results for $p$Pb and Pb+Pb collisions.  (The $p_T$ 
dependence is stronger in the Pb+Pb result.)
(b) We show the dependence of $R_{AB}$ on the various production channels
in Pb+Pb collisions at $\sqrt{s}=5.5$~TeV.
}
\end{figure}

Better determinations of the $\Upsilon$ matrix elements are required in
order to make more accurate predictions of the NRQCD
$\Upsilon$-production rates at the LHC. As is shown in the
$\Upsilon$-polarization analysis in Ref.~\cite{Braaten:2000gw}, some
theoretical predictions have quite large uncertainties even at Tevatron
energies, owing to our poor knowledge of the matrix elements. However,
as is shown in Fig.~\ref{lee.fig4}, the ratio $R_{AB}$ is still a good
measure of the effect of shadowing on $\Upsilon$ production.  The ratio
is independent of the $\Upsilon$ state and is quite similar to the
$J/\psi$ ratio in Fig.~\ref{lee.fig2}.  The shadowing effect in Pb+Pb
interactions may be somewhat less for the $\Upsilon$ at $p_T \approx 5$
GeV than for the $J/\psi$, but the difference is small.  Note also, from
Fig.~\ref{lee.fig4}(b), that $R_{AB}$ is essentially independent of the
matrix elements and is, therefore, largely unaffected by their
uncertainties.

\section*{Acknowledgments}

Work by G.~T.~Bodwin and J.~Lee in the High Energy Physics Division at
Argonne National Laboratory is supported by the U.~S.~Department of
Energy, Division of High Energy Physics, under Contract
No.~W-31-109-ENG-38. Work by R.~Vogt is supported in part by the
Division of Nuclear Physics of the Office of High Energy and Nuclear
Physics of the U.~S.~Department of Energy under Contract
No.~DE-AC-03-76SF00098.

\end{document}